\documentclass[a4paper,aps,prl,twocolumn,showpacs,preprintnumbers,amsmath,amssymb]{revtex4}
\usepackage{bbm}
\usepackage{amsmath}
\usepackage{verbatim}
\usepackage{graphicx}
\usepackage{graphics}
\usepackage{natbib}
\usepackage{textcomp}
\usepackage{color}
\usepackage{hyperref}
\usepackage{times}

\newcommand{\xc}{x_{\rm c}}
\newcommand{\pc}{p_{\rm c}}
\newcommand{\Xm}{X_{\rm m}}
\newcommand{\Pm}{P_{\rm m}}
\newcommand{\Xa}{X_{\rm a}}
\newcommand{\Pa}{P_{\rm a}}
\newcommand{\Xaa}{X_{\rm a2}}
\newcommand{\Paa}{P_{\rm a2}}
\newcommand{\omm}{\omega_{\rm m}}
\newcommand{\omc}{\omega_{\rm c}}

\newcommand{\aad}{a_{\rm a}^{\dagger}}
\newcommand{\ac}{a_{\rm c}^{\phantom\dagger}}
\newcommand{\acd}{a_{\rm c}^{\dagger}}

\newcommand{\gm}{\gamma_{\rm m}}
\newcommand{\gc}{\gamma_{\rm c}}

\newcommand{\xinc}{x_{\rm in}}
\newcommand{\pinc}{p_{\rm in}}
\newcommand{\xoutc}{x_{\rm out}}
\newcommand{\poutc}{p_{\rm out}}

\newcommand{\xina}{x{\scriptstyle'}_{\!\rm in}}
\newcommand{\pina}{p{\scriptstyle'}_{\!\rm in}}
\newcommand{\xouta}{x{\scriptstyle'}_{\!\!\rm out}}
\newcommand{\pouta}{p{\scriptstyle'}_{\!\!\rm out}}

\newcommand{\poutcos}{p_{\rm out}^{\rm cos}}
\newcommand{\poutsin}{p_{\rm out}^{\rm sin}}
\newcommand{\pincos}{p_{\rm in}^{\rm cos}}
\newcommand{\pinsin}{p_{\rm in}^{\rm sin}}

\newcommand{\ket}[1]{|#1\rangle}
\newcommand{\bra}[1]{\langle #1|}
\newcommand{\mean}[1]{\langle #1\rangle}

\newcommand{\Dt}{\frac{d}{dt}}

\newcommand{\Nph}{N_{\rm ph}}
\newcommand{\Nat}{N_{\rm at}}

\newcommand{\nin}{\bar{n}_{\rm i}}
\newcommand{\nth}{\bar{n}_{\rm th}}

\newcommand{\Qm}{Q_{\rm m}}

\definecolor{grey}{rgb}{0.8,0.8,0.8}

\begin{document}

\title{Establishing EPR-channels between Nanomechanics and Atomic Ensembles}

\author{K. Hammerer$^{1,2}$, M. Aspelmeyer$^{2}$, E.S. Polzik$^3$, P. Zoller$^{1,2}$}

\affiliation{ $^1$Institut f\"ur Theoretische Physik, Universit\"at Innsbruck, A-6020 Innsbruck, Austria \\
$^2$Institute for Quantum Optics and Quantum Information (IQOQI) of the
Austrian Academy of Sciences, A-6020 Innsbruck and A-1090 Vienna, Austria\\
$^3$Niels Bohr Institute, QUANTOP, Danish Research Foundation Center for Quantum Optics, Blegdamsvej, DK-2100 Copenhagen, Denmark}

\begin{abstract}
We suggest to interface nanomechanical systems via an optical quantum bus to atomic ensembles, for which means of high precision state preparation, manipulation and measurement are available. This allows in particular for a Quantum Non-Demolition Bell measurement, projecting the coupled system, atomic ensemble - nanomechanical resonator, into an entangled EPR-state. The entanglement is observable even for nanoresonators initially well above their ground states and can be utilized for teleportation of states from an atomic ensemble to the mechanical system.
\end{abstract}

\pacs{03.67.-a,03.65.Ud}

\maketitle

Opto- and electro-nanomechanical systems \cite{Cite:Reviews}, representing cold high-Q oscillators coupled to optical cavities or electrical circuits, are rapidly approaching the regime where quantum aspects are important \cite{Cite:NanoresExp,Cite:NanoresCooling,Cite:NanoresEntanglement,Brown2007,Thompson2008}. It remains one of the key challenges of nanomechanics to develop both the tools for preparing and manipulating quantum states as superposition and entangled states, and to implement quantum measurements. Motivated by the remarkable achievements in the quantum control of atomic ensembles \cite{Cite:BeamSplitter,Cite:DownConversion,Kuzmich2000,Cite:Copenhagen1,Cite:Copenhagen2,Cite:Faraday}, which allow for high-fidelity preparation, readout, and laser manipulation of atomic states as long-lived quantum memory, we propose a quantum interface between atomic ensembles and opto-mechanical systems, where light plays the role of a quantum bus. This effort should be seen in the context of developing hybrid systems and interfaces where the goal is to combine the advantages of solid state and atomic systems in compatible experimental setups.

Our goal is the creation of robust Einstein-Podolsky-Rosen (EPR) type of entanglement \cite{Cite:ContVar} between collective spin variables of the atomic medium and a nanomechanical resonator. EPR entanglement is a key resource in many quantum information protocols, and in particular enables the manipulation and transfer of quantum states, e.g. by quantum teleportation. EPR states involve two systems, each described by a pair of continuous variables, say $[\Xm,\Pm]=i$ and $[\Xa,\Pa]=i$, and exhibit a reduced so-called EPR variance in their correlations
\begin{align}\label{Eq:EPRvariance}
  \Delta_{\rm EPR}=\Delta(\Xm+\Xa)^2+\Delta(\Pm-\Pa)^2<2.
\end{align}
The value of 2 refers to uncorrelated systems in their ground states and any value below 2 proves entanglement \cite{Cite:EntCrit}. In the present case canonical variables $\Xm,\,\Pm$ refer to (dimensionless) position and momentum of the mechanical oscillator, while $\Xa,\,\Pa$ describe collective spin excitations in an atomic ensemble as follows: The fully polarized state of an ensemble of atoms, each with two stable ground states $\ket\pm$, is identified with the ground state of a harmonic oscillator, $\ket{+\cdots+}=\ket{0}$ and excited states are given by $\ket{1}=\aad\ket{0}$ etc., where $\aad=\Nat^{-1/2}J_+$, $\Nat$ is the number of atoms and $J_+=\sum_i\ket-_i\bra+$ is a collective step up operator. Accordingly, canonical operators correspond to scaled collective spin components $\Xa=(\Nat/2)^{-1/2}J_x$ and $\Pa=(\Nat/2)^{-1/2}J_y$. This is an excellent approximation for the high degree of polarization routinely achieved in current experiments \cite{Cite:BeamSplitter,Cite:DownConversion,Kuzmich2000,Cite:Copenhagen1,Cite:Copenhagen2}.

The method to generate EPR correlations suggested here is based on a Quantum Non--Demolition (QND) measurement \cite{Cite:QND} of commuting EPR observables $\Xm+\Xa$ and $\Pm-\Pa$. Using light as a meter system, as  was demonstrated for two atomic ensembles by Julsgaard et al. \cite{Cite:Copenhagen1}, the QND measurement projects the hybrid nanomechanical-atomic system into a state with reduced variances in the EPR observables that obey \eqref{Eq:EPRvariance}. The method relies on the fact that both systems can be coupled to a cavity mode $[\xc,\pc]=i$ via structurally similar Hamiltonians
\begin{align}
H_{\rm mc}&=\frac{\omm}{2}\left(\Xm^2+\Pm^2\right)+g\Xm\xc,\label{Eq:Hmc}\\
H_{\rm ac}&=\frac{\Omega}{2}\left(\Xa^2+\Pa^2\right)+G\Xa\xc,\label{Eq:Hac}
\end{align}
where $\omm$ is the mechanical frequency, $\Omega$ denotes the energy splitting of the two atomic levels and $g$ and $G$ are coupling constants. The physical basis of \eqref{Eq:Hmc} and \eqref{Eq:Hac} are, respectively, radiation pressure \cite{Cite:Reviews} and Faraday interaction \cite{Hammerer2008} as will be detailed further below. The basic principle underlying the QND measurement is best explained by assuming for the moment that both systems are coupled to the same cavity mode, and that we tune $G=g$ and $\Omega=-\omm$. The Hamiltonian is then the sum of \eqref{Eq:Hmc} and \eqref{Eq:Hac}, and in an interaction picture with respect to the harmonic oscillator terms the resulting Hamiltonian is
\begin{align*}
H_I=g\left[\cos(\omm t)\xc(\Xm+\Xa)+\sin(\omm t)\xc(\Pm-\Pa)\right].
\end{align*}
Evidently, the relevant EPR observables are conserved quantities and -- given the cavity decay happens fast on a time scale of $g$ -- cosine and sine components of light coupled out of the cavity will read out these EPR observables, making them detectable in a homodyne measurement of light. Note that it is vital to choose $\Omega=-\omm$, i.e. to let the atomic ensemble realize a harmonic oscillator with {\em negative mass}, in order to have {\it commuting} EPR observables appearing in $H_I$ and, therefore, a QND interaction, a situation whose realization is not obvious for two nanomechanical oscillators.

\begin{figure}
\includegraphics[width=8.6cm]{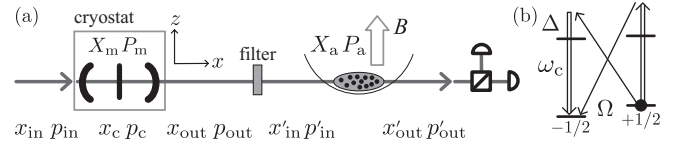}
\caption{(a) Schematic of setup: A pulse of light described by canonical operators $(\xinc,\pinc)$ interacts first with a micromechanical oscillator $(\Xm,\Pm)$, realized by e.g. a membrane coupled to a Fabry-Perot cavity mode $(\xc,\pc)$. Light leaving the cavity $(\xoutc,\poutc)$ is phase modulated by the membrane. A filter turns these phase- into polarization-modulation. Subsequently the pulse $(\xinc',\pinc')$ interacts with the collective spin of an atomic ensemble, effectively described by canonical operators $\Xa,\Pa$, and Larmor precessing in a magnetic field $B$. Light $(\xoutc',\poutc')$ leaving the  ensemble is subject to homodyne detection. (b) Atomic level scheme with quantization axis along $z$: Atoms are prepared in the state $+1/2$ of a spin $j=1/2$ ground state, Zeeman split by $\Omega$.
Light propagates along $x$ with a classical field polarized along $z$ driving the $\pm 1/2\leftrightarrow \pm 1/2$ transitions (double arrows) and the $y$ polarized quantum sidebands entangled with the nanosystem coupling to the $\pm 1/2\leftrightarrow \mp1/2$ transitions (thin arrows).}\label{Fig:Setup}
\end{figure}

As will be detailed below, the QND measurement and EPR state preparation can actually be achieved in a {\em cascaded quantum system} according to  Fig.~\ref{Fig:Setup}, where the output light of the laser driven optomechanical system is coupled to an atomic ensemble at a (possibly) distant location, followed by a homodyne measurement. This setup has three remarkable features: First, this establishes {\em distant} EPR correlations, as is familiar from quantum communications with continuous variables systems, and avoids the requirement of holding the cloud of atoms close to the opto-nanomechanical system. Second, the present protocol remarkably does \emph{not} require ground state cooling of the nanomechanical resonator. In the limit of strong coupling a QND measurement realizes a projective von Neumann measurement, collapsing the systems into a pure, entangled state irrespective of initial conditions. Surprisingly, this takes effect already for a QND measurement of moderate and well feasible strength. Third, this setup also provides measurement and verification of a reduced EPR variance by simply repeating the protocol.

Recently two proposals for hybrid quantum systems involving atoms and nano-systems were put forward \cite{Treutlein2007, Genes2008}. Our proposal is distinctly different from the one by Treutlein et al.~\cite{Treutlein2007}, which suggests direct coupling of a Bose Einstein condensate to a magnetic island on a cantilever. Entanglement between short-lived electronically excited states of an atomic ensemble and a nanomechanical system, {\it both} being placed inside a cavity, has been very recently studied theoretically by Genes et al.~\cite{Genes2008}.

The detailed  treatment of light propagation, losses, as well as realistic conditions for the matching of time scales will be the main content of the remaining parts of this letter. We start with a brief derivation of the fundamental Hamiltonians \eqref{Eq:Hmc} and \eqref{Eq:Hac}, which are both well established models in their respective fields, see \cite{Cite:NanoresExp,Cite:NanoresCooling,Cite:NanoresEntanglement,Brown2007,Thompson2008} and \cite{Cite:Copenhagen1,Cite:Copenhagen2,Cite:Faraday} respectively. In the opto-mechanical system the fundamental interaction is based on radiation pressure \cite{Law1995,Thompson2008} described by $V=g_0\acd\ac\Xm$, where $g_0=(x_0/L)\omc$ and $x_0$ is the mechanical oscillator ground state spread, $L$ the cavity length and $\omc$ its frequency. If the cavity is driven by a resonant pulse of duration $\tau\gg1/\gc$, where $\gc$ is the cavity decay rate, and power $P=\Nph\hbar\omc/\tau$ containing a total number $\Nph$ of photons, a steady state amplitude $\alpha=\sqrt{\Nph/\tau\gc}$ builds up and the dynamics can be linearized giving an effective interaction $V_{\rm eff}=g\xc\Xm$ as in \eqref{Eq:Hmc}, where $\xc,\,\pc$ describe fluctuations of the cavity field, see e.g. \cite{Cite:NanoresEntanglement}. From Hamiltonian \eqref{Eq:Hmc} the evolution is given by
\begin{align*}
\dot x_{\rm c}&=-\gc \xc\!-\!\sqrt{2\gc}\,\xinc, &
\dot p_{\rm c}&=-\gc \pc\!-\!\sqrt{2\gc}\,\pinc-g \Xm,
\end{align*}
where $[\xinc(t),\pinc(t')]=i\delta(t-t')$ denotes vacuum noise. Assuming $\gc\gg g,\,\omm$ we adiabatically eliminate the cavity mode and arrive at the cavity input-output relations \cite{Gardiner2000}
\begin{align}\label{Eq:inoutc}
  \xoutc&=-\xinc, &
  \poutc&=-\pinc-g\sqrt{2/\gc}\Xm,
\end{align}
These expressions refer to field quadratures slowly varying around $\omc$. According to the second equation in \eqref{Eq:inoutc}, the phase quadrature variance will be above shot noise due to correlations to the mechanical quadrature $\Xm$ carried by Stokes- and Antistokes-sideband photons at $\omc\pm\omm$.

As indicated in Fig.\ref{Fig:Setup}a the beam leaving the cavity interacts with an ensemble of $\Nat$ atoms in free space, with a relevant level scheme shown in Fig.\ref{Fig:Setup}b. For simplicity we will give a derivation of \eqref{Eq:Hac} for atoms inside a low Finesse cavity and derive from it input-output relations, as was done above. The same input-output relations can be shown to hold in free space \cite{Cite:Faraday} and provide an excellent description of experiments \cite{Cite:Copenhagen1,Cite:Copenhagen2}. The level configuration in Fig.\ref{Fig:Setup}b gives rise to the Hamiltonian $H\propto J_xS_z$, where $S_z=-i(a^\dagger_ya^{\phantom{\dagger}}_z-a^\dagger_za^{\phantom{\dagger}}_y)/2$ is a Stokes vector defined for two polarization modes $[a_i^{\phantom\dagger},a^\dagger_j]=\delta_{i,j}\,(i,j=y,z)$. This so-called Faraday interaction describes mutual polarization rotation of the atomic spins and the cavity field. If the $z$-polarized cavity mode is coherently driven, such that $\mean{a_z}=i|\alpha|$, one can approximate $S_z\simeq|\alpha|(a^{\phantom{\dagger}}_y+a^\dagger_y)/2\propto \xc$. Using canonical variables $\Xa,\Pa$ for collective spin components $J_x,J_y$ as explained above results in an interaction $H\propto \Xa\xc$. For non-degenerate ground states there will in addition be a free Hamiltonian $H_0=\Omega J_z=-(\Omega/2)(\Xa^2+\Pa^2)$, where the minus sign is due to the fact that atoms are pumped to the energetically higher lying state cf. Fig.\ref{Fig:Setup}b. Overall, we arrive at a Hamiltonian \eqref{Eq:Hac} with $\Omega\rightarrow-\Omega$. Adiabatic elimination of the cavity mode, just as for the mechanical system, will yield input-output relations
\begin{align*}
  \xouta&=-\xina, & 
  \pouta&=-\pina-\kappa\sqrt{2/\tau}\Xa.
\end{align*}
In free space $\tau$ is the pulse length and $\kappa^2=(\sigma\Gamma/A\Delta)^2\Nat\Nph$, where $\Gamma$ is the spontaneous decay rate, $\Delta$ the detuning, $\sigma$ the scattering- and $A$ the beam cross section \cite{Cite:Faraday}.

We now assume that the cavity output provides the input to the light-atoms interaction such that $\xina=-\xoutc$ and $\pina=-\poutc$. This requires that the coherent pulse at frequency $\omc$ is rotated in polarization by $90^\circ$ and phase shifted by $\pi/2$ relative to its sideband components at $\omc\pm\omm$, which can be achieved by separating the optical carrier and the sidebands with an auxiliary cavity and then performing the required rotations and shifts. Note that because the sidebands will be measured by homodyning with the carrier, an well feasible extinction ratio for the carrier at the percent level is sufficient \cite{Neergaard2006}. We now assume a matching of time scales by requiring
\begin{align}\label{Eq:Numbermatch}
\kappa/\sqrt{\tau}=g/\sqrt{\gc},
\end{align}
which can be fulfilled experimentally as indicated below. Under these conditions the overall input-output relations become
\begin{align}\label{Eq:TotalInOut}
  \xouta&=-\xinc, &
  \pouta&=-\pinc-\kappa\sqrt{2/\tau}(\Xm+\Xa).
\end{align}

In order to achieve a QND measurement of EPR variables $\Xm+\Xa$ and $\Pm-\Pa$, these variables have to be free of back action of light and light has to read out both. This is indeed the case. From the discussion above it follows straight forwardly that the mechanical oscillator evolves as,
\begin{equation*}
  \dot X_{\rm m}\!=\!\omm\Pm,\,\,\,
  \dot P_{\rm m}\!=\!-\omm\Xm\!-\!g\xc\!=\!-\omm\Xm\!+\!g\sqrt{2/\gc}\xinc,
\end{equation*}
where in the last equality the cavity mode was again adiabatically eliminated. In these equations we neglect the decay of the oscillator, which is justified if the whole protocol happens on a time scale $\sim \tau$ such that \mbox{$\tau\gm\nth\ll 1$}. Here $\gm$ is the mechanical damping rate and $\nth=k_BT/\hbar\omm$ is the mean occupation in thermal equilibrium at temperature $T$. If this condition is met, decay can be treated perturbatively, as will be done below. Transverse atomic spin components evolve as
\begin{align*}
  \dot X_{\rm a}&=-\Omega\Pa,& 
  \dot P_{\rm a}&=\Omega\Xa+\frac{\kappa}{\sqrt{\tau}}\xina=\Omega\Xa+\kappa\sqrt{\frac{2}{\tau}}\,\xinc.
\end{align*}
Decoherence due to spontaneous emission can be kept small for sufficient optical depth \cite{Cite:Faraday} and will be included perturbatively further below. Using again condition \eqref{Eq:Numbermatch} and taking in addition $\omm=\Omega$, we finally arrive at
\begin{align*}
  \textstyle{\Dt}(\Xm\!+\!\Xa)&\!=\!\Omega(\Pm\!-\!\Pa),&\!\!
  \textstyle{\Dt}(\Pm\!-\!\Pa)&\!=\!-\Omega(\Xm\!+\!\Xa).
\end{align*}
In this closed set of equations of motion for commuting EPR observables $\Xm+\Xa$ and $\Pm+\Pa$ back action effects of light cancel out by interference. This establishes the QND character of the present interactions.

The QND signal lies essentially in the Fourier components at frequency $\Omega$ of the in-quadrature component $\pouta$. For the normalized observables $\poutcos=-\sqrt{2/\tau}\int_0^\tau{\rm d}t\cos(\Omega t)\pouta(t)$ and $\poutsin=-\sqrt{2/\tau}\int_0^\tau{\rm d}t\sin(\Omega t)\pouta(t)$ one readily derives from \eqref{Eq:TotalInOut} the input-output relations,
\begin{equation}\label{Eq:BellQND}
  \poutcos\!=\!\pincos\!+\kappa(\Xm\!+\Xa)_{\rm in},\,\,\,
  \poutsin\!=\!\pinsin\!+\kappa(\Pm\!-\Pa)_{\rm in},
\end{equation}
with appropriate definitions for the in-components $x^{\rm in}_{\rm cos(sin)}$. We assume here $\Omega \tau\gg1$ such that cosine and sine components can be taken as independent variables.

A measurement of $x_{\rm out}^{\rm cos(sin)}$ leaves the mechanical resonator and the collective spin in a state with reduced EPR variance \eqref{Eq:EPRvariance}, \emph{conditioned} on the respective measurement results $\xi_{\rm cos(sin)}$ of $x_{\rm out}^{\rm cos(sin)}$. An unconditionally reduced variance can be achieved by a feedback operation on the atomic spin, e.g. via fast rf pulses causing appropriate tilt of the spin, generating a displacement $\Xa\rightarrow\Xa-g\xi_{\rm cos},\,\Pa\rightarrow\Pa-g\xi_{\rm sin}$ with a suitable gain $g$. In the ensemble average this generates a state, whose statistics is described by the input-output relations \cite{Hammerer2005}
\begin{align}
  (\Xm+\Xa)_{\rm out}&=(\Xm\!+\Xa)_{\rm in}-g\poutcos\label{Eq:Feedback}\\
  &=(1-g\kappa)(\Xm\!+\Xa)_{\rm in}-g\pincos,\nonumber\\
  (\Pm-\Pa)_{\rm out}&=(1-g\kappa)(\Pm\!-\Pa)_{\rm in}-g\pinsin.\nonumber
\end{align}
Our aim is to minimize this variance with respect to the feedback gain $g$. As initial state of the systems we assume vacuum for light modes, the ground (fully polarized) state for the collective spin and an initial thermal occupation $\nin$ for the mechanical resonator. In thermal equilibrium $\nin=\nth$, but in principle $\nin$ can be reduced by initial laser cooling \cite{Cite:NanoresExp,Cite:NanoresCooling}. For optimal gain the minimal EPR variance is
\begin{align}\label{Eq:EPR}
  \Delta_{\rm EPR}=\frac{2}{\frac{1}{(1+\nin)}+2\kappa^2},
\end{align}
which is the main result of this paper.

According to Eq.~\eqref{Eq:EPRvariance} this is an entangled state if the right hand side of \eqref{Eq:EPR} falls below 2. As $2[(1+\nin)^{-1}+2\kappa^2]^{-1}\leq\kappa^{-2}$ \emph{there is no fundamental limit on observable entanglement due to initial thermal occupation $\nin$ of the mechanical system}. Thus, even for moderate values of $\kappa\gtrsim 0.5$, achievable as outlined below, the present protocol produces an entangled state independent of the initial thermal occupation of the nanomechanical resonator.

We turn to the discussion of losses and decoherence. The dominant impairing effects are (i) number mismatch in Eq.\eqref{Eq:Numbermatch}, (ii) loss of light, detection inefficiency and spontaneous emission in light-atom interactions, and (iii) thermalization of the mechanical oscillator. Ad (i), it is straight forward to derive that a nonzero mismatch $\epsilon=(\kappa-\lambda\sqrt{\gc\tau})/(\kappa+\lambda\sqrt{\gc\tau})$ will give rise to additional terms in the variance of EPR variables \eqref{Eq:EPR} scaling in leading order as $[\epsilon\kappa(\nin+2)]^2$. For $\kappa\gtrsim 1$ a mismatch of $\epsilon\simeq 1/10\nin$ is tolerable. This poses a practical limit to the initial thermal occupation of the nanomechanical resonator. Effects due to processes (ii) and (iii) can be treated perturbatively as linear losses, as we exemplify for damping of the resonator: During the interaction the state of the resonator will undergo damping at a rate $\gm$ and provided $\gm\tau\ll1$ e.g. Eq.\eqref{Eq:Feedback} will generalize to
\[(\Xm+\Xa)_{\rm out}=(\sqrt{1-\gm\tau}\Xm+\Xa)_{\rm in}+\sqrt{\gm\tau}f_{\Xm}-g\poutcos,\]
where $f_{\Xm}$ is a Langevin operator for thermal noise, $\mean{f_{\Xm}^2}=(\nth+1)$. The variance will thus receive an additional term $\gm\tau(\nth+1)$, such that we have to require $\tau\ll1/\gm\nth=\Qm\hbar/k_BT$ for a quality factor $\Qm=\omm/\gm$. In a similar vein one can treat losses (ii), which have the advantageous property that the corresponding noise is, to a good approximation, vacuum noise. That is, a photon loss by a fraction $\varepsilon$ will cause the final EPR variance to be $\Delta_{\rm EPR}\rightarrow(1-\varepsilon)\Delta_{\rm EPR}+2\varepsilon$. This will reduce but not remove the entanglement created by this protocol.

The suggested protocol can be realized with current technology. We consider two possible setups, in which the nanomechanical resonator is used either as one of the mirrors of a Fabry-Perot cavity \cite{Cite:NanoresExp} or as a dispersive element in a rigid cavity \cite{Thompson2008}. Assuming that $\kappa\simeq1$ and that condition~\eqref{Eq:Numbermatch} can be matched within an error of $\epsilon=0.01$ we need $\nin\lesssim30$. This can be achieved for high $\omm$ at dilution refrigerator temperatures, cf. Naik et al. \cite{Cite:NanoresExp} or, in case of lower $\omm$ or higher bath temperatures, by applying additional laser cooling. As an example for the two cases we assume a moving micromirror with $\omm/2\pi=5~{\rm MHz}$, mass $m=10^{-12}~{\rm kg}$ and quality factor $\Qm\geq5\times10^5$ operated at $T=0.2~{\rm K}$ (resulting in $\nin\approx850$, which requires modest laser cooling by a factor of 30), and a small (dispersively coupled) nanomembrane with $\omm/2\pi=30~{\rm MHz}$, $m=10^{-14}~{\rm kg}$ and $\Qm\geq10^5$ operated at $T=0.04~{\rm K}$ ($\nin\approx 30$). Mechanical quality and temperature limit the interaction time to $\tau\ll20~\mu{\rm s}$, which is in principle sufficient for entanglement of room temperature atoms and certainly enough in the case of cold atoms. For the laser-cooled micromirror \eqref{Eq:Numbermatch} can be achieved with a finesse $F=4500$ and power $P=100~\mu{\rm W}$. Adiabatic elimination of the cavity mode finally poses an upper bound $L\leq300~\mu{\rm m}$ on the cavity length. For the nanomembrane a modest finesse $F=1100$ is already sufficient at a pump power $P=100~\mu{\rm W}$ and cavity length $L\leq250~\mu{\rm m}$.

Finally, the generated entanglement can serve as a basis for teleporting quantum states of a collective spin onto a nanomechanical system. The protocol proceeds in three steps. First, an entangled state characterized by Eq.~\eqref{Eq:EPR} is created and a second, additional atomic ensemble is prepared in a coherent state with amplitudes $\langle\Xaa\rangle,\,\langle\Paa\rangle$. Second, following the approach of \cite{Cite:Copenhagen1} a QND Bell measurement of $(\Xaa+\Xa)$ and $(\Paa-\Pa)$ is performed on the two atomic ensembles. Third, the measurement result is used in a feedback on the mechanical system, via e.g. piezo-electric or radiation pressure displacement. This completes the teleportation and generates a state,
\begin{align*}
  \Xm^{\rm fin}&=\Xm+g[\pincos\!+\kappa_{\rm QND}(\Xaa+\Xa)]=\Xm+\Xa+\Xaa,\\
  \Pm^{\rm fin}&=\Pm+g[\pinsin\!+\kappa_{\rm QND}(\Paa-\Pa)]=\Pm-\Pa+\Paa.
\end{align*}
Here $\kappa_{\rm QND}$ and $g$ denote strength of QND interaction and feedback gain in the Bell measurement on the two atomic ensembles. The second equalities of both lines are valid in the asymptotic limit $\kappa_{\rm QND}\rightarrow\infty,\,g\rightarrow 0$ while $\kappa_{\rm QND}g=1$, which essentially requires a large optical depth \cite{Hammerer2008,Cite:Faraday}. Amplitudes are thus transmitted correctly, $\langle\Xm\rangle=\langle\Xaa\rangle$ and $\langle\Pm\rangle=\langle\Paa\rangle$, and the amount of added noise is given by $\Delta_{\rm EPR}/2$ in Eq.~\eqref{Eq:EPR}, as e.g. $\Delta(\Xm^{\rm fin})^2=\Delta(\Xaa)^2+\Delta(\Xm+\Xa)^2$ and equivalently for $\Delta(\Pm^{\rm fin})^2$. For $\kappa\simeq 1$ this is approximately one unit of vacuum noise in each variable, corresponding to a fidelity of $2/3$. We note that this implies the intriguing possibility to cool a mechanical resonator by teleporting the ground state onto it. More details of the proposed protocols will be presented in a forthcoming publication.

We acknowledge support by the FWF under SFB F15, doctoral program W1210 and project P15939, by the Foundational Questions Institute (grant RFP1-06-14), and by EU projects QAP, COMPAS, EuroSQIP and EMALI.

%
%
%
%

\end{document}